\documentclass[showpacs, pra, aps, twocolumn,10pt]{revtex4}
\usepackage{mathrsfs}
\usepackage{array}
\usepackage{amsmath}
\usepackage{graphicx}
\usepackage{amstext}
\usepackage{bm}
\usepackage{amsfonts}
\hfuzz=\maxdimen
\begin{document}
\title{Steady-state one-way Einstein-Podolsky-Rosen steering in optomechanical interfaces}
\author{Huatang Tan}
\email{tanhuatang@phy.ccnu.edu.cn}
\affiliation{Department of Physics, Huazhong Normal University, Wuhan 430079, China}
\author{Xincheng Zhang}
\affiliation{Department of Physics, Huazhong Normal University, Wuhan 430079, China}
\author{Gaoxiang Li}
\affiliation{Department of Physics, Huazhong Normal University, Wuhan 430079, China}

\begin{abstract}
Einstein-Podolsky-Rosen (EPR) steering is a form of quantum correlations and its intrinsic asymmetry makes it distinct from entanglement and Bell nonlocality. We propose here a scheme for realizing one-way Gaussian steering of two electromagnetic fields mediated by a mechanical oscillator. We reveal that the steady-state one-way steering of the intracavity and output fields is obtainable with different cavity losses or strong mechanical damping. The conditions for achieving this asymmetric steering are found, and it shows that the steering is robust against thermal mechanical fluctuations. The present scheme can realize hybrid microwave-optical asymmetric steering by optoelectromechanics. In addition, our results are generic and can also be applied to other three-mode parametrically-coupled bosonic systems.
\end{abstract}

\maketitle

\section{Introduction}
Steering was initially introduced by Schr\"{o}dinger \cite{Str} in response to the famous EPR paradox proposed by Einstein, Podolsky, and Rosen in 1935 \cite{EPR}. The paradox describes that two remote observers Alice and Bob share a pair of entangled particles and one observer, say Alice, can prepare the state of Bob's particle via different types of measurements on her own particle. Steering was termed as Alice's ability to nonlocally control Bob's state via local measurements.

Recently, steering has been revisted and rigorously formulated in Refs. \cite{wiseman01,wiseman02,wiseman03}. The violation of a local hidden-state model for sceptical Bob demonstrates steering from Alice to Bob.  It shows that Bell-nonlocal states violating Bell inequality \cite{Bell01,Bell02} are subset of the steerable states which in turn are a subset of the inseparable states. Steering embodies a kind of quantum correlations intermediate between entanglement and Bell nonlocality. Any demonstration of EPR paradox is also a demonstration of steering and vice versa \cite{wiseman03}. EPR paradox was firstly realized by Ou et al \cite{ou} and steering has recently been experimentally realized in different systems \cite{bowen,wiseman04,cfli}. Besides being of fundamental interest, quantum steering is useful for quantum information such as quantum cryptography \cite{qucry}.

Inherently distinct from entanglement and Bell nonlocality, steering is intrinsically asymmetric between the two observers. That is, the roles played in steering by Alice and Bob are not exchangeable. Very interestingly, recent theoretical and experimental works have verified there exists asymmetric steering, i.e. one-way steering, which allows Alice's steering the state of Bob's particle but the reverse Bob-to-Alice steering is impossible. This one-way steering reflects the asymmetry of quantum correlations. One-way Gaussian steering has been experimentally achieved by controlling unequal losses of two entangled beams \cite{schnabel}, and theoretical studies have also revealed this asymmetric steering in several systems of continuous and discrete variables \cite{olsen01,olsen02,qyhe01,qyhe02,bowles}.

In this paper, we propose a scheme for realizing one-way Gaussian steering of two electromagnetic fields by optomechanics with continuous pumps. In the past decade, considerable progress has been made in the field of quantum optomechanics \cite{opm}. Quantum ground-state cooling of mechanical oscillators \cite{cool}, mechanically-induced squeezing \cite{squ1}, and optomechanical entanglement \cite{omen} have been achieved. It makes optomechanical interfaces a very promising platform for demonstrating various quantum phenomena. Our system consists of two driven electromagnetic cavities mediated by a mechanical oscillator. Photon entanglement in such optomechanical interfaces has been studied in detail \cite{ybc,tht,vitali,tian,ydw,xiao}. Here we focus on the steerability and asymmetry of the photon correlations. The conditions for achieving one-way steering in different directions for the cases of weak and strong mechanical damping are identified. Our scheme can realize hybrid microwave-optical steering in optoelectromechanical interface.

\section{Model}
We consider a double cavity optomechanical system in which two separate cavity fields of frequencies $\omega_{cj}~(j=1,2)$ are mediated by a mechanical oscillator at frequency $\omega_m$. The cavity fields, driven by coherent fields of frequencies $\omega_{lj}$, can be optical modes \cite{chan}, microwave modes \cite{massel}, or both \cite{regal,tianl} (see Fig.1 (a)). In particular, a recent experiment has realized the reversible transfer between microwave and optical photons with a mechanical element. This optoelectromechanical interface may allow for quantum information processing with light at different wavelengths by exploiting  microwave-optical quantum correlations \cite{vitali}. Strong photon nonlinearity can also be achieved in such a setup, as studied in Ref.\cite{lv}. For strong driving fields, the linearized Hamiltonian of the system reads
\begin{align}
\hat H_1=\omega_m \hat b^\dag \hat b+\sum_{j=1}^2\Big[\Delta_j\hat a_j^\dag \hat a_j+g_j(\hat a_j^\dag+\hat a_j)(\hat b+\hat b^\dag)\Big],
\label{ham1}
\end{align}
where the bosonic operators $\hat a_j$ and $\hat b$ describe the cavity and mechanical modes, respectively, with $\Delta_j$ and $g_j$ the effective cavity-drive detunings and optomechanical couplings. The couplings $g_j$ are controllable via changing the drive strengths.
\begin{figure}
\centerline{\scalebox{0.30}{\includegraphics{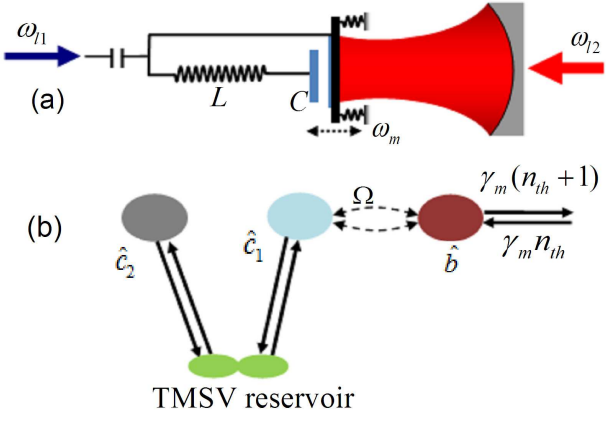}}}
 \caption{(a) The schematic plot of double cavity optomechanics in which two separate electromagnetic fields (e.g. microwave and optical field) are mediated by a mechanical oscillator vibrating at frequency $\omega_m$. The two cavity fields are respectively driven on the red and blue sidebands of the driving fields. (b) After the squeezing transformation of Eq.(\ref{st}), the composite mode $\hat c_2$ is decoupled to the mechanical mode $\hat b$ which interacts with the composite mode $\hat c_1$ via a beam-splitter-like interaction and meanwhile the two composite modes are coupled to an effective reservoir in two-mode squeezed vacuum. The mechanical damping drives the steady two-mode cavity field states to be asymmetric.}
\end{figure}

We consider that the cavity field 1 and 2 are resonant with the red and blue sidebands of the driving fields, respectively,  i.e., $\Delta_1=-\omega_m$ and $\Delta_2=\omega_m$. In an interaction picture with respect to $\hat H_0=\sum_j\Delta_j \hat a_j^\dag \hat a_j$ and under the rotation-wave approximation (RWA), the Hamiltonian of Eq.(\ref{ham1}) becomes into
\begin{align}
\hat H_2=g_1 (\hat a_1\hat b+\hat a_1^\dag\hat b^\dag)+g_2(\hat a_2\hat b^\dag+\hat a_2^\dag\hat b).
\label{ham2}
\end{align}
The parametric downconversion can bring about the optomechanical entanglement,
while the upconversion induces quantum-state transfer between the cavity field 2 and the mechanical oscillator.
These two combined processes lead to the entanglement between the two cavity fields.
The above Hamiltonian represents a typical three-mode parametric interaction and can be realized in some other bosonic systems  \cite{gx, parkins, olsen03}, e.g. atomic ensembles coupled to optical fields \cite{hammerer}, apart from the optomechanical interfaces. Our results are therefore generic and applicable to these systems.

The system's operators are governed by
\begin{align}
&\partial _t\hat a_1^\dag=-\kappa_1\hat a_1^\dag+ig_1\hat b+\sqrt{2\kappa_1}\hat a_1^{\rm in\dag}(t),\nonumber\\
&\partial _t\hat a_2=-\kappa_2\hat a_2-ig_2\hat b+\sqrt{2\kappa_2}\hat a_2^{\rm in}(t),\label{lan1}\\
&\partial _t\hat b=-\gamma_m\hat b-ig_1\hat a_1^\dag-ig_2\hat a_2 +\sqrt{2\gamma_m}\hat b^{\rm in}(t),\nonumber
\end{align}
where $\kappa_j$ and $\gamma_m$ are the dissipation rates of the cavity and mechanical modes. The noise operators $\hat a_j^{\rm in}(t)$ and $\hat b^{\rm in}(t)$ satisfy nonzero correlations $\langle \hat a_j^{\rm in}(t)\hat a_{j'}^{\rm in\dag}(t)\rangle=\delta_{jj'}\delta(t-t')$, $\langle \hat b^{\rm in\dag}(t)\hat b^{\rm in}(t')\rangle=\bar{n}_{\rm th}\delta(t-t')$,  and $\langle \hat b^{\rm in}(t)\hat b^{\rm in\dag}(t')\rangle=(\bar{n}_{\rm th}+1)\delta(t-t')$, where $\bar{n}_{\rm th}=(e^{\hbar\omega_m/k_BT}-1)^{-1}$, $T$ the temperature and $k_B$ the Boltzmann constant. Note that the Hamiltonian of Eq.(\ref{ham2}) under the RWA is only valid under the condition
\begin{align}
\omega_m\gg \{g_j,~\kappa_j,~\gamma_m \bar{n}_{\rm th}\}.
\label{crwa}
\end{align}
With the Routh-Hurwitz criterion, the stability condition of Eq.(\ref{lan1}) can be found to be
\begin{align}
&(\kappa_2+\gamma_m)[(\kappa_1+\kappa_2)(\kappa_1+\gamma_m)+g_2^2]>(\kappa_1+\gamma_m)g_1^2,\nonumber\\
&\kappa_1g_2^2-\kappa_2g_1^2+\gamma_m\kappa_1\kappa_2>0.\label{stcond}
\end{align}

\section{EPR-Steering criteria}
For the quadrature operators $\hat X_j=\hat a_j+\hat a_j^\dag$ and $\hat Y_j=-i(\hat a_j-\hat a_j^\dag)$, the Heisenberg uncertainty relations  are $V(\hat X_j)V(\hat Y_j)\ge1$, where the variances $V(\hat {\mathcal O})=\langle \hat {\mathcal O}^2\rangle-\langle \hat {\mathcal O}\rangle^2$ for $ \mathcal{\hat O}=(\hat X_j, \hat Y_j)$. According to Refs.\cite{wiseman03, reid}, the EPR paradox and steering of bipartite Gaussian states are achievable on Gaussian measurements when
\begin{align}
S_{12}=V_{\rm inf}(\hat X_1)V_{\rm inf}(\hat Y_1)<1,
\label{str12}
\end{align}
or
\begin{align}
S_{21}=V_{\rm inf}(\hat X_2)V_{\rm inf}(\hat Y_2)<1.
\label{str21}
\end{align}
The inferred variances $V_{\rm inf}(\hat {\mathcal O}_j)$ are $
V_{\rm inf}\big[\hat {\mathcal O}_{1(2)}\big]=V\big[\hat {\mathcal O}_{1(2)}\big]-V^2(\hat {\mathcal O}_1, \hat {\mathcal O}_2)/V\big[\hat {\mathcal O}_{2(1)}\big]$. The condition $S_{12}<1$ ($S_{21}<1$) means the steerability from the cavity field 2 (1) to the field $1$ (2). One-way steering occurs when only one of the above two inequalities holds.

Specifically, for our system the steady average values $\langle \hat a_j^2\rangle_{ss}=0$ and $\langle \hat a_1 \hat a_2^\dag\rangle_{ss}=0$ (see Appendix \ref{ap1}). The steering criteria of Eqs.(\ref{str12}) and (\ref{str21}) then reduce respectively to
\begin{align}
\big|\langle \hat a_1\hat a_2\rangle_{ss}\big|>\sqrt{\langle \hat a_1^\dag \hat a_1\rangle_{ss}(\langle \hat a_2^\dag \hat a_2\rangle_{ss}+1/2)},
\label{eqstr12}
\end{align}
and
\begin{align}
\big|\langle \hat a_1 \hat a_2\rangle_{ss}\big|>\sqrt{\langle \hat a_2^\dag \hat a_2\rangle_{ss}(\langle \hat a_1^\dag \hat a_1\rangle_{ss}+1/2)}.
\label{eqstr21}
\end{align}
While the entanglement between the cavity fields measured by logarithmic negativity \cite{eln} requires (see Appendix \ref{ap2})
\begin{align}
\big|\langle \hat a_1 \hat a_2\rangle_{ss}\big|>\sqrt{\langle \hat a_1^\dag \hat a_1\rangle_{ss}\langle\hat a_2^\dag \hat a_2\rangle_{ss}}.
\label{en1}
\end{align}
We see that nonclassical correlations between the two fields are necessary for the entanglement
and steering and moreover stronger nonclassical correlations are required for
achieving the steering than that for the entanglement. It therefore exemplifies that steerable states are strictly inseparable but not necessarily vice versa.

\section{Steering of intracavity fields}
\subsection{Weak mechanical damping regime $(\gamma_m\ll\kappa_j)$}
\begin{figure}[t]
\centerline{\scalebox{0.5}{\includegraphics{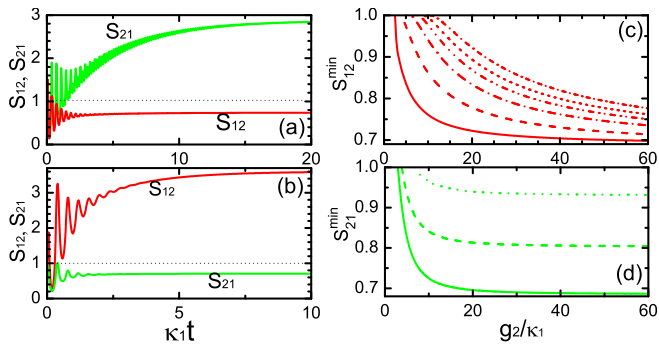}}}
 \caption{ $S_{12}$ and $S_{21}$ \emph{versus} time for $\kappa_2=0.4\kappa_1$ and $g_1=10\kappa_1$ in (a) and $\kappa_2=2.4\kappa_1$ and $g_1=12\kappa_1$ in (b), with $g_2=20\kappa_1$, $\gamma_m=0.01\kappa_1$ and $\bar{n}_{\rm th}=0$.  The values of $\kappa_2$ and $g_1$ are chosen such that $S_{12}$ in (a) and $S_{21}$ in (b) achieve their minima. (c) and (d):  Minimized $S_{12}^{\rm min}$ and $S_{21}^{\rm min}$ with respect to $g_1/\kappa_1$ and $\kappa_2/\kappa_1$
\emph{versus} $g_2/\kappa_1$, with the values of $\kappa_2/\kappa_1$ are the same as in (a) and (b), respectively. From top to bottom, $\bar{n}_{\rm th}=1000,~700,~500,~300,~100,~0$ in (c) and $\bar{n}_{\rm th}=40,~20,~0$ in (d).}
\label{case1}
\end{figure}
At first, we consider the case that $g_j>\kappa_j\gg\gamma_m$ such that we can temporarily neglect the mechanical damping at zero temperature for simplicity. Then, the steering conditions $S_{12}<1$ and $S_{21}<1$ for the steady cavity states reduce respectively to
\begin{align}
(\kappa_1-\kappa_2)(\kappa_2g_2^2-\kappa_1g_1^2)>\kappa_1\kappa_2(\kappa_1+\kappa_2)^2,
\label{st122}
\end{align}
and
\begin{align}
(\kappa_2-\kappa_1)(\kappa_2g_2^2-\kappa_1g_1^2)>\kappa_1\kappa_2(\kappa_1+\kappa_2)^2,
\label{st212}
\end{align}
which are incompatible. The condition for the steady-state entanglement reduces to
\begin{align}
(\kappa_2g_2^2-\kappa_1g_1^2)>0.
\label{en2}
\end{align}
For the same cavity losses ($\kappa_1=\kappa_2$), both Eq.(\ref{st122}) and (\ref{st212}) can not be held and therefore the steady entangled cavity-field states are definitely not steerable at zero mechanical damping. For different cavity loss rates, the directions of one-way steering depends on the ratio of $\kappa_2/\kappa_1$. We see that the one-way steering from the cavity field 2 to the cavity field 1 ($S_{12}<1$ and $S_{21}>1$) may be achieved
when $\kappa_2>\kappa_1$, whereas the reverse one-way steering may occur
when $\kappa_2<\kappa_1$. This result is plotted in Fig.\ref{case1} (a) and (b)
by considering $\gamma_m=10^{-2}\kappa_1$, and it shows that steady-state one-way steering in two ways can be obtained, apart from the transient two-way steering ($S_{12}<1$ and $S_{21}<1$). We thus conclude that in the steady-state regime only the cavity field with larger dissipation rate can be steered by the other one. This is because that in the absence of the mechanical damping, the field under larger dissipation has smaller steady-state mean photon number and also smaller quantum fluctuations \big[$V(\hat {\mathcal O}_j)=\langle \hat a_j^\dag \hat a_j\rangle$+1/2\big]. This field is therefore more easier steered by the other one (since it has larger fluctuations and thus smaller inferred variances of the steered field).

Fig. \ref{case1} (c) and (d) plot the dependence of the minimized $S_{12}^{\rm min}$ and $S_{21}^{\rm min}$ (maximized steering), with respect to the couplings $g_1/\kappa_1$ and $\kappa_2/\kappa_1$ in the steady-state regime, on $g_2/\kappa_1$ at nonzero temperature. We see that the one-way steering from the field 2 to 1 is much more robust against thermal fluctuations than the reverse one. This is due to that the thermal input leads to the much more enhancement of the mean photon number $\langle \hat a_2^\dag\hat a_2\rangle_{ss}$ than that of $\langle \hat a_1^\dag \hat a_1\rangle_{ss}$ via the beam-splitter interaction of the cavity field $\hat a_2$ with the mechanical mode coupled to the thermal reservoir. Therefore, even for the environment with large thermal phonon number, the one-way field 2-to-1 steering can also be achievable. This shows that the present scheme can realize asymmetric steering without precooling the mechanical oscillator to its ground state.

\subsection{Strong mechanical damping regime $(\gamma_m\gg\kappa_j)$}
\begin{figure}
\centerline{\scalebox{0.63}{\includegraphics{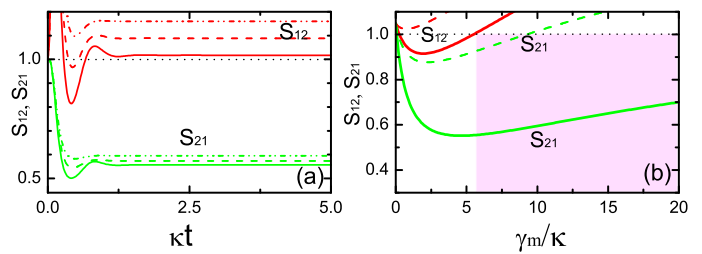}}}
 \caption{(a) $S_{12}$ and $S_{21}$ \emph{versus} time at zero temperature for $\gamma_m=6\kappa$ (solid curves), $\gamma_m=8\kappa$ (dashed curves), and $\gamma_m=10\kappa$ (dashed-dotted-dotted curves). (b) The steady-state $S_{12}$ and $S_{21}$ \emph{versus} $\gamma_m/\kappa$ for $\bar{n}_{\rm th}=0$ (solid curves) and $\bar{n}_{\rm th}=0.3$ (dashed curves). The other parameters $g_1=6\kappa$ and $g_2=10\kappa$.}
 \label{case2}
\end{figure}
We next study the role played by the strong mechanical damping in steering. For simplicity, we assume the cavity dissipation rates $\kappa_j=\kappa$. In this case and at zero temperature, the steady-state entanglement is always present and the steering condition $S_{21}<1$ reduces to
\begin{align}
\gamma_m/\kappa>1/\big[(\Omega/2\kappa)^2-1\big],
\label{str213}
\end{align}
with $\Omega=\sqrt{g_2^2-g_1^2}$. It shows that for $\Omega\gg2\kappa$, weak mechanical damping can still lead to the cavity field 1-to-2 steering. Meanwhile, the steering condition $S_{12}<1$ becomes approximately into
\begin{align}
\gamma_m/\kappa<\big(g_2\sqrt{\Omega^2-8\kappa^2}-\Omega^2\big)/2\kappa^2,
\label{str123}
\end{align}
for $\gamma_m\gg\kappa$. When $\Omega\gg2\kappa$, the right hand of Eq.(\ref{str123}) is larger than that of Eq.(\ref{str213}) and therefore the one-way steering from the cavity field 1 to 2 can be obtained for $\gamma_m$ violating Eq.(\ref{str123}). This is shown in Fig.\ref{case2} (a) and (b). We see that strong mechanical damping in vacuum leads to the steady-state cavity field 1-to-2 one-way steering and its strength is impaired by thermal mechanical noise. The maximum obtainable steering is larger than that in the case of weak mechanical damping in Fig. \ref{case1}. In addition, we find that in this regime of strong mechanical damping, the reverse one-way steering from the cavity field 2 to 1 is unobtainable (see below for reason) and strong two-way steering can be achieved by minimizing $S_{12}$ and $S_{21}$ with respect to $g_j/\kappa$ (see Fig. \ref{fig6} in the Appendix).
\begin{figure}
\centerline{\scalebox{0.55}{\includegraphics{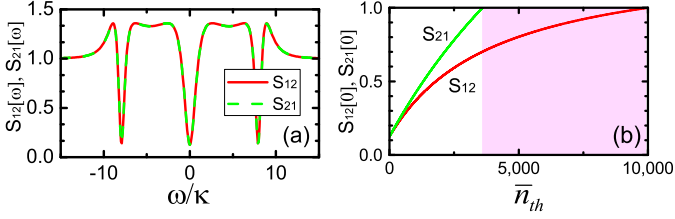}}}
 \caption{(a) The spectra of $S_{12}[\omega]$ and $S_{21}[\omega]$ for $g_1=6\kappa$, $g_2=10\kappa$, $\gamma_m=0.01\kappa$, and $\bar{n}_{\rm th}=0$. (b) $S_{12}[0]$ and $S_{21}[0]$ \emph{versus} $\bar{n}_{\rm th}$. }
\label{case3}
\end{figure}

To show the mechanical damping can lead to the asymmetric cavity states, we perform a transformation
\begin{align}
\hat c_j=\hat S(r)\hat a_j \hat S(-r),
\label{st}
\end{align}
where the two-mode squeezing operator $\hat S(r)=\exp\big[r(\hat a_1^\dag\hat a_2^\dag-a_1\hat a_2)\big]$ and
$r=\tanh^{-1}(g_2/g_1)$, giving $\hat c_1=\sinh r \hat a_1^\dag+\cosh r\hat a_2$ and $\hat c_2=\cosh r \hat a_1+\sinh r\hat a_2^\dag$. In terms of the operators $\hat c_j$, the equations of Eq.(\ref{lan1}) become into
$\partial _t\hat c_1=-\kappa\hat c_1-i\Omega\hat b+\sqrt{2\kappa}\hat c_1^{\rm in}(t)$,
$\partial _t\hat c_2=-\kappa\hat c_2+\sqrt{2\kappa}\hat c_2^{\rm in}(t)$, and
$\partial _t\hat b=-\gamma_m\hat b-i\Omega \hat c_1 +\sqrt{2\gamma_m}\hat b^{\rm in}(t)$, where
the noise operators $\hat c_j^{\rm in}(t)$ satisfy the nonzero correlations $\langle \hat c_j^{\rm in\dag}(t)\hat c_j^{\rm in}(t')\rangle=\sinh^2r\delta(t-t')$, $\langle \hat c_j^{\rm in}(t)\hat c_j^{\rm in\dag}(t')\rangle=\cosh^2r\delta(t-t')$, and $\langle \hat c_1^{\rm in}(t)\hat c_2^{\rm in}(t')\rangle=\sinh r\cosh r\delta(t-t')$. We see that the composite mode $\hat c_2$ is decoupled to the mechanical oscillator interacting with the mode $\hat c_1$ via $\hat H_{c1b}=\Omega \hat c_1\hat b^\dag+h.c$ and the two modes $\hat c_j$ are coupled to a two-mode squeezed vacuum reservoir (see Fig.1 (b)). The symmetry of the cavity field states depends on that of $\hat c_1$ and $\hat c_2$ modes. For zero mechanical damping, the the two composite modes reduce to reservoir's state and thus symmetric. However, at finite mechanical damping in vacuum the mode $\hat c_1$ is cooled down and we have $\langle \hat a_1^\dag\hat a_1\rangle_{ss}-\langle \hat a_2^\dag \hat a_2\rangle_{ss}=\langle \hat c_2^\dag\hat c_2\rangle_{ss}-\langle \hat c_1^\dag \hat c_1\rangle_{ss}>0$, which results in the violation of Eq.(\ref{eqstr12}) more easily than that of Eq.(\ref{eqstr21}) and thus the one-way steering of the field 2 by the field 1.

We note that a large mechanical damping rate can be obtained by using a low-quality mechanical oscillator with high resonant frequency. Alternately, it can also be achieved by weakly coupling a high-quality mechanical oscillator to a bad electromagnetic cavity to induce an optical heating for the mechanical oscillator, as analyzed in detail in Ref.\cite{nunnen}.

\section{Output steering spectra}
\begin{figure}
\centerline{\scalebox{0.55}{\includegraphics{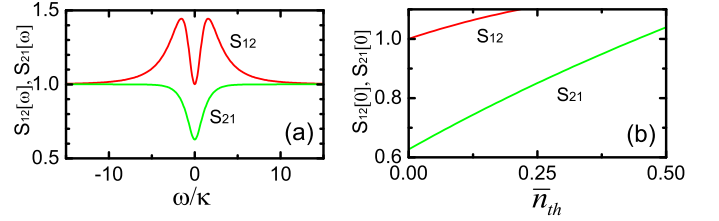}}}
 \caption{(a) The spectra of $S_{12}[\omega]$ and $S_{21}[\omega]$ for $g_1=2\kappa$, $g_2=3\kappa$, $\gamma_m=9\kappa$, and $\bar{n}_{\rm th}=0$. (b) $S_{12}[0]$ and $S_{21}[0]$ \emph{versus} $\bar{n}_{\rm th}$.}
 \label{case4}
 \end{figure}
We finally consider the steering spectra of the output fields which are utilized for measurements and applications. By performing the Fourier transformation $\hat O(t)=\int_{-\infty}^{\infty}e^{-i\omega
t}\hat O[\omega]d\omega/\sqrt{2\pi}$ on Eq.(\ref{lan1}) and using the input-output relations $ \hat a_j^{\rm out}=\sqrt{2\kappa_j}\hat a_j-\hat a_j^{\rm in}$, we have
\begin{align}
\hat a_1^{\rm out}[\omega]=&M_{11}\hat a_1^{\rm in}[\omega] +M_{12}\hat a_2^{\rm in\dag}[-\omega]+M_{1b} \hat b^{\rm in\dag}[-\omega],\nonumber \\
\hat a_2^{\rm out}[\omega]=& -M_{12} \hat a_1^{\rm in\dag}[-\omega] +M_{22}\hat a_2^{\rm in}[\omega] +M_{2b}\hat b^{\rm in}[\omega].
\end{align}
The expressions for $M_{jj'}$ and $M_{jb}$ and the calculation of the steering spectra $S_{12}[\omega]$ and $S_{21}[\omega]$ are given in Appendix \ref{ap4}.

At zero mechanical damping, we have $\big\langle\hat a_1^{\rm out\dag}[\omega]a_1^{\rm out}[\omega]\big\rangle=\big\langle\hat a_2^{\rm out\dag}[\omega]a_2^{\rm out}[\omega]\big\rangle=\big|m_{12}^2\big|$, which means that the output states at frequencies $\omega_{cj}+\omega$ are symmetric, independent of the ratio $\kappa_2/\kappa_1$. Therefore, unlike the intracavity situation, one-way spectral steering can impossibly be achieved when $\gamma_m=0$, even for the unbalanced cavity losses. The symmetric steering spectra are plotted in Fig.\ref{case3} (a). It shows that strong two-way steering can be obtained around the frequencies
\begin{align}
\omega=0,~~\omega=\pm\sqrt{\Omega^2-\kappa^2},
\end{align}
at which the cavity fields are strongly excited. Nevertheless, for weak mechanical damping the one-way steering from the output field 2 to the field 1 at cavity resonances is obtainable via thermalizing the mechanical oscillator, as shown in Fig\ref{case3} (b), since the steering in this direction is more robust against thermal noise than the reverse steering, as similarly in Figs.\ref{case1} (c) and (d). This one-way steering at cavity resonances ($S_{12}[0]<1$ and $S_{21}[0]>1$)
requires
\begin{align}
g_1^2/\kappa\gamma_m< \bar{n}_{\rm th}<g_2^2/\kappa\gamma_m-1.
\end{align}
Hence, the one-way steering from the output field 2 to 1 can still be achieved even for the large number of thermal phonons when $g_j^2/\kappa\gamma_m\gg1$. It shows again that this directional one-way steering is much more robust to thermal noise than the reverse one, as exhibited in Fig.(\ref{case1}).

By increasing the mechanical damping rate which enters the strong damping regime, we plot the steering in Fig.\ref{case4}. Interestingly, we see that the one-way steering from the output field 1 to the field 2 can be achievable over all frequencies. At zero temperature the condition $S_{21}[0]<1$ always holds while $S_{12}[0]>1$ requires the mechanical damping rate
\begin{align}
\gamma_m/\kappa>g_2^2/\kappa^2,
\end{align}
for which the one-way steering in this direction is achievable.

\section{Conclusion}
In summary, we propose a scheme for realizing one-way steering of two electromagnetic fields by double cavity optomechanics. The two cavity fields are mediated by a mechanical oscillator and driven respectively by a red and a blue detuned strong coherent fields. We show that asymmetric steering can be achieved for unequal cavity losses or strong mechanical damping in the regime of steady states. The conditions for achieving one-way steering of the intracavity and output fields are found. The asymmetric steering may be useful in quantum communication and information. Besides optoelectromechanical interfaces, our results are also applicable to other three-mode parametrically-coupled bosonic systems. Further work will consider tripartite steering in the present system.

\section*{Acknowledgments}
This work is supported by the National Natural Science Foundation of China (Grant Nos.~11274134, 61275123, and 11474119), and the National Basic Research Programm of China (2012CB921602).

\appendix
\begin{widetext}
\section{The steady-state solution of the system}
\label{ap1}
The equations (\ref{lan1}) can be rewritten into the simple form
\begin{align}
\frac{d}{dt} \psi=\textbf{A}\psi+\sqrt{2\textbf{K}}\psi_{\rm in}(t),
\label{lan2}
\end{align}
where $\psi=(\hat a_1, \hat a_1^\dag, \hat a_2, \hat a_2^\dag,\hat b, \hat b^\dag)^T$, $\textbf{K}=diag(\kappa_1, \kappa_1, \kappa_2,\kappa_2,\gamma_m,\gamma_m)$, and $\psi_{\rm in}=(\hat a_1^{\rm in}, \hat a_1^{\rm in\dag},\hat a_2^{\rm in},\hat a_2^{\rm in\dag},\hat b^{\rm in},\hat b^{\rm in \dag})^T$. The matrix
\begin{align}
\textbf{A}=\left(
  \begin{array}{cccccc}
    -\kappa_1& 0 & 0 & 0 & 0 & -ig_1 \\
    0 & -\kappa_1 & 0 & 0 & ig_1 & 0 \\
    0 & 0 &  -\kappa_2& 0 & -ig_2 & 0 \\
    0 & 0 & 0 & -\kappa_2 & 0 & ig_2 \\
    0 & -ig_1 & -ig_2 & 0& -\gamma_m & 0 \\
    ig_1 & 0 & 0 & ig_2 & 0 & -\gamma_m \\
  \end{array}
\right).
\end{align}
From Eq.(\ref{lan2}) the second-order moments $\Phi=\langle\psi\psi^T\rangle$ satisfy
\begin{align}
\frac{d}{dt}\Phi=\textbf{A}\Phi+\Phi \textbf{A}^T+2\textbf{K}\textbf{D},
\label{inside}
\end{align}
where $\textbf{D}=diag(\textbf{D}_1,\textbf{D}_2,\textbf{D}_3)$, with the entries
$\textbf{D}_{1,2}=\left(\begin{array}{cccc}0 & 1 \\ 0 & 0\\ \end{array}\right)$ and $\textbf{D}_3=\left(\begin{array}{cccc}0 & \bar{n}_{\rm th}+1 \\ \bar{n}_{th} & 0\\ \end{array}\right)$. In the steady-state regime, we have
\begin{subequations}
\begin{align}
\langle \hat a_1^\dag\hat a_1\rangle_{ss}&=\frac{\kappa_2(\kappa_1+\kappa_2+\gamma_m)g_1^2g_2^2+\gamma_m (n_{\rm th}+1)\big[\kappa_1g_2^2-\kappa_2g_1^2+\kappa_2(\kappa_1+\kappa_2)(\kappa_2+\gamma_m)\big]g_1^2}{(\kappa_1g_2^2-\kappa_2g_1^2+\gamma_m\kappa_1\kappa_2) \big[(\kappa_2+\gamma_m)g_2^2-(\kappa_1+\gamma_m)g_1^2+(\kappa_1+\kappa_2)(\kappa_1+\gamma_m)(\kappa_2+\gamma_m)\big]},\\
\langle \hat a_2^\dag\hat a_2\rangle_{ss}&=\frac{\kappa_1(\kappa_1+\kappa_2+\gamma_m)g_1^2g_2^2+\gamma_m n_{\rm th}\big[\kappa_1g_2^2-\kappa_2g_1^2+\kappa_1(\kappa_1+\kappa_2)(\kappa_1+\gamma_m)\big]g_2^2}{(\kappa_1g_2^2-\kappa_2g_1^2+\gamma_m\kappa_1\kappa_2) \big[(\kappa_2+\gamma_m)g_2^2-(\kappa_1+\gamma_m)g_1^2+(\kappa_1+\kappa_2)(\kappa_1+\gamma_m)(\kappa_2+\gamma_m)\big]},\\
\langle \hat a_1\hat a_2\rangle_{ss}&=-\frac{\kappa_1g_1g_2\big[\kappa_2g_1^2+(\kappa_2+\gamma_m)(g_2^2+\kappa_2\gamma_m)\big]+\gamma_m n_{\rm th}\big[\kappa_1g_2^2-\kappa_2g_1^2+\kappa_1\kappa_2(\kappa_1+\kappa_2+2\gamma_m)\big]}{(\kappa_1g_2^2-\kappa_2g_1^2+\gamma_m\kappa_1\kappa_2) \big[(\kappa_2+\gamma_m)g_2^2-(\kappa_1+\gamma_m)g_1^2+(\kappa_1+\kappa_2)(\kappa_1+\gamma_m)(\kappa_2+\gamma_m)\big]}.
\end{align}
\end{subequations}

\section{The derivation of the entanglement condition in Eq.(\ref{eqen})}
\label{ap2}
The correlation matrix $\sigma$ of the steady two-mode cavity field states, defined as $\sigma_{ij}=\langle \xi_i\xi_j+\xi_j\xi_i\rangle/2$ and $\xi=(\hat X_1, \hat Y_1, \hat X_2,\hat Y_2)$, can be obtained as
\begin{align}
\sigma=\left(
    \begin{array}{cccc}
    n_1& 0 & c & 0\\
    0 & n_1 & 0 & -c\\
    c & 0 & n_2& 0\\
    0 & -c & 0 & n_2\\
  \end{array}
  \right)
\end{align}
with $n_j=\langle \hat a_j^\dag\hat a_j\rangle_{ss}+1/2$ and $c=\langle \hat a_1\hat a_2\rangle_{ss}$. Here, the quadrature operators are scaled by 1/2 with respect to their definition in the text. By expressing the
correlation matrix $\sigma$ in terms of three $2\times2$ matrices $\sigma_1$, $\sigma_2$, and $\sigma_3$ as $
\sigma=\begin{pmatrix} \sigma_1 & \sigma_3 \\
\sigma_3^T & \sigma_2\end{pmatrix}$, the logarithmic negativity $E_N$ is defined as
\begin{align}
E_N=\text {max}[0, -\ln(2\lambda)],
\end{align}
where $\lambda=2^{-1/2}\sqrt{\Sigma(\sigma)-\sqrt{\Sigma(\sigma)-4\text{det}\sigma}}$ and $\Sigma(\sigma)=\text {det}\sigma_1+\text {det} \sigma_2-2\text{det} \sigma_3$. The entanglement thus occurs for $\lambda<1/2$ which is equivalent to
\begin{align}
\big[c^2-(n_1+1/2)(n_2+1/2)\big]\big[c^2-(n_1-1/2)(n_2-1/2)\big]<0.
\end{align}
Since the positivity of the two-mode cavity-field states requires that $c^2<(n_1+1/2)(n_2+1/2)$, the above entanglement condition reduces to
\begin{align}
\big|\langle \hat a_1 \hat a_2\rangle_{ss}\big|>\sqrt{\langle \hat a_1^\dag \hat a_1\rangle_{ss}\langle\hat a_2^\dag \hat a_2\rangle_{ss}}.
\end{align}


\begin{figure}[t]
\centerline{\scalebox{0.5}{\includegraphics{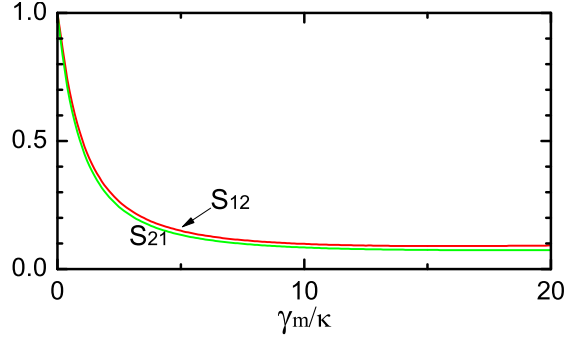}}}
 \caption{The dependence of the minimized $S_{12}$ and $S_{21}$, with respect to $g_1/\kappa$ and $g_2/\kappa$, in the steady-state regime on the $\gamma_m/\kappa$, for $n_{\rm th}=0$.}
 \label{fig6}
\end{figure}

\section{The calculation of the steering spectra}
\label{ap4}
By performing the Fourier transformation $\hat O(t)=\int_{-\infty}^{\infty}e^{-i\omega t}\hat O[\omega]d\omega/\sqrt{2\pi}$ on Eq.(\ref{lan1}), we have
\begin{align}
-i\omega \hat a_1^\dag [-\omega]&=-\kappa_1\hat a_1^\dag[-\omega]+ig_1\hat b[\omega]+\sqrt{2\kappa_1}\hat a_1^{\rm in\dag }[-\omega],\nonumber\\
-i\omega \hat a_2 [\omega]&=-\kappa_2\hat a_1^\dag[\omega]+ig_2\hat b[\omega]+\sqrt{2\kappa_2}\hat a_2^{\rm in\dag }[\omega],\\
-i\omega \hat b [\omega]&=-\gamma_m\hat b[\omega]-ig_1\hat a_1^\dag[-\omega]-ig_2\hat a_2[\omega]+\sqrt{2\gamma_m}\hat b^{\rm in}[\omega].\nonumber
\end{align}
With the input-output relations $ \hat a_j^{\rm out}[\omega]=\sqrt{2\kappa_j}\hat a_j[\omega]-\hat a_j^{\rm in}[\omega]$, we can express the output fields, in terms of the input ones, as
\begin{align}
\hat a_1^{\rm out}[\omega]=&M_{11}\hat a_1^{\rm in}[\omega] +M_{12}\hat a_2^{\rm in\dag}[-\omega]+M_{1b} \hat b^{\rm in\dag}[-\omega],\nonumber \\
\hat a_2^{\rm out}[\omega]=& -M_{12} \hat a_1^{\rm in\dag}[-\omega] +M_{22}\hat a_2^{\rm in}[\omega] +M_{2b}\hat b^{\rm in}[\omega].
\end{align}
where
\begin{align}
M_{11}[\omega]&=\frac{(\kappa_1+i\omega)g_2^2+(\kappa_2-i\omega)g_1^2+(\kappa_1+i\omega)(\kappa_2-i\omega)(\gamma_m-i\omega)}{
(\kappa_1-i\omega)g_2^2-(\kappa_2-i\omega)g_1^2+(\kappa_1-i\omega)(\kappa_2-i\omega) (\gamma_m-i\omega)},\nonumber\\
M_{12}[\omega]&=\frac{2\sqrt{\kappa_1\kappa_2}g_1g_2}{
(\kappa_1-i\omega)g_2^2-(\kappa_2-i\omega)g_1^2+(\kappa_1-i\omega)(\kappa_2-i\omega) (\gamma_m-i\omega)},\nonumber\\
M_{1b}[\omega]&=\frac{-2i\sqrt{\kappa_1\gamma_m}g_1(\kappa_2-i\omega)}{
(\kappa_1-i\omega)g_2^2-(\kappa_2-i\omega)g_1^2+(\kappa_1-i\omega)(\kappa_2-i\omega) (\gamma_m-i\omega)},\nonumber\\
M_{22}[\omega]&=-\frac{(\kappa_1+i\omega)g_2^2+(\kappa_2+i\omega)g_1^2-(\kappa_1-i\omega)(\kappa_2+i\omega)(\gamma_m-i\omega)}{
(\kappa_1-i\omega)g_2^2-(\kappa_2-i\omega)g_1^2+(\kappa_1-i\omega)(\kappa_2-i\omega) (\gamma_m-i\omega)},\nonumber\\
M_{2b}[\omega]&=\frac{-2i\sqrt{\kappa_2\gamma_m}g_2(\kappa_1-i\omega)}{
(\kappa_1-i\omega)g_2^2-(\kappa_2-i\omega)g_1^2+(\kappa_1-i\omega)(\kappa_2-i\omega) (\gamma_m-i\omega)}.\nonumber
\end{align}
With the spectral definitions of the quadratures $\hat X_j^{\rm out}[\omega]=\hat a_j^{\rm out}[\omega]
+\hat a_j^{\rm out \dag} [-\omega]$ and $\hat Y_j^{\rm out}[\omega]=-i\hat a_j^{\rm out}[\omega]+i\hat a_j^{\rm out \dag} [-\omega]$, we have
\begin{align}
\langle \hat X_1^2[\omega]\rangle=\langle Y_1^2[\omega]\rangle&=\big|M_{11}[\omega]\big|^2+\big|M_{12}[-\omega]\big|^2+\big|M_{1b}[-\omega]\big|^2(\bar{n}_{\rm th}+1)+\big|M_{1b}[\omega]\big|^2\bar{n}_{\rm th},\\
\langle \hat X_2^2[\omega]\rangle=\langle Y_2^2[\omega]\rangle&=\big|M_{22}[\omega]\big|^2+\big|M_{12}[-\omega]\big|^2+\big|M_{2b}[\omega]\big|^2(n_{\rm th}+1)+\big|M_{2b}[-\omega]\big|^2\bar{n}_{\rm th},\\
\langle \hat X_1[\omega]\hat X_2[\omega]\rangle&=-\langle \hat Y_1[\omega]\hat Y_2[\omega]\rangle=-M_{11}[\omega]M_{12}[-\omega]+M_{12}^*[-\omega]M_{22}^*[\omega]\nonumber\\
&~~~~~~~~~~~~~~~~~~~~~~~~~+M_{1b}^*[-\omega]M_{2b}^*[\omega](\bar{n}_{\rm th}+1)+M_{1b}[\omega]M_{2b}[-\omega]\bar{n}_{\rm th}.
\end{align}
\end{widetext}

\end{document}